\documentclass{sig-alt-full}

\begin{document}

\conferenceinfo{DOLAP'08,} {October 30, 2008, Napa Valley,
California, USA.}
\CopyrightYear{2008}
\crdata{978-1-60558-250-4/08/10}

\title{Data Mining-based Fragmentation \\of XML Data Warehouses}

\numberofauthors{2}
\author{
\alignauthor
Hadj Mahboubi\\
       \affaddr{University of Lyon (ERIC Lyon 2)}\\
       \affaddr{5 avenue Pierre Mend\`{e}s-France}\\
       \affaddr{69676 Bron Cedex -- France}\\
       \email{hadj.mahboubi@eric.univ-lyon2.fr}
\alignauthor
J\'{e}r\^{o}me Darmont\\
       \affaddr{University of Lyon (ERIC Lyon 2)}\\
       \affaddr{5 avenue Pierre Mend\`{e}s-France}\\
       \affaddr{69676 Bron Cedex -- France}\\
       \email{jerome.darmont@univ-lyon2.fr}
}

\maketitle

\begin{abstract}
With the multiplication of XML data sources, many XML data warehouse
models have been proposed to handle data heterogeneity and
complexity in a way relational data warehouses fail to achieve.
However, XML-native database systems currently suffer from limited
performances, both in terms of manageable data volume and response
time. Fragmentation helps address both these issues. Derived
horizontal fragmentation is typically used in relational data
warehouses and can definitely be adapted to the XML context.
However, the number of fragments produced by classical algorithms is
difficult to control. In this paper, we propose the use of a
k-means-based fragmentation approach that allows to master the
number of fragments through its $k$ parameter. We experimentally
compare its efficiency to classical derived horizontal fragmentation
algorithms adapted to XML data warehouses and show its superiority.
\end{abstract}




\section{Introduction}

XML data sources that are pertinent for decision-support are becoming increasingly common with XML becoming a standard for representing complex business data \cite{BeyerCCOP05}. However, they bear specificities (e.g., heterogeneous number and order of dimensions or complex measures in facts, ragged dimension hierarchies, etc.) that would be intricate to handle in a relational environment. Hence, many efforts toward XML data warehousing have been achieved in the past few years \cite{BoussaidMCA06,Pokorny02,ZhangWLZ05}, as well as efforts for extending the XQuery language with near On-Line Analytical Processing (OLAP) capabilities, such as advanced grouping and aggregation features \cite{BeyerCCOP05,datax08,WiwatwattanaJLS07}.

In this context, performance is a critical issue, since XML-native
database systems currently suffer from limited performances, both in
terms of manageable data volume and response time for complex
analytical queries. These issues are typical of data warehouses and
can be addressed by fragmentation. Fragmentation consists in
splitting a data set into several fragments such that their
combination yields the original warehouse without loss nor
information addition. Fragmentation can subsequently lead to
distribute the target data warehouse, e.g., on a data
grid~\cite{CostaF06}. In the relational context, derived horizontal
fragmentation is acknowledged as best-suited to data warehouses,
because it takes decision-support query requirements into
consideration and avoids computing unnecessary join operations
\cite{BellatrecheB05}. Several approaches have also been proposed
for XML data fragmentation, but they do not take multidimensional
architectures (i.e., star-like schemas) into account.

In derived horizontal fragmentation, dimensions' primary horizontal fragmentation is crucial. In the relational context, two major algorithms address this issue: the predicate construction \cite{NoamanB99} and the affinity-based \cite{NavatheKR95} strategies. However, both are automatic and the number of fragments is not known in advance, while it is crucial to master it, especially since distributing $M$ fragments over $N$ nodes with $M > N$ can become an issue in itself. Hence, we propose in this paper the use of a k-means-based fragmentation approach that allows to control the number of fragments through its $k$ parameter. Our idea, which adapts a proposal from the object-oriented domain \cite{Darabant04} to XML warehouses, is to cluster workload query predicates to produce primary horizontal dimension fragments, with one fragment corresponding to one cluster of predicates. Primary fragmentation is then derived on facts. Queries including given predicates are executed over the corresponding fragments only, instead of the whole warehouse, and thus run faster. 

The remainder of this paper is organized as follows. We first
discuss existing research related to relational data warehouse, XML
data and data mining-based fragmentation in
Section~\ref{sec:RelatedWork}. Then, we present our k-means-based
XML data warehouse fragmentation approach in
Section~\ref{sec:KMeansBasedFragmentation}. We experimentally
compare its efficiency to classical derived horizontal fragmentation
algorithms adapted to XML data warehouses and show its superiority
in Section~\ref{sec:ExperimentalPerformanceStudy}. Finally, we
conclude this paper and present future research directions in
Section~\ref{sec:Conclusion}.

\pagebreak

\section{Related Work}
\label{sec:RelatedWork}

\subsection{Fragmentation Definition}

There are three fragmentation types in the relational context~\cite{KoreichiB97}: vertical fragmentation, horizontal fragmentation
and hybrid fragmentation.

Vertical fragmentation splits a relation $R$ into subrelations that
are projections of $R$ with respect to a subset of attributes. It
consists in grouping together attributes that are frequently
accessed by queries. Vertical fragments are built by projection. The
original relation is reconstructed by joining the fragments.

Horizontal fragmentation divides a relation into subsets of tuples
using query predicates. It reduces query processing costs by
minimizing the number of irrelevant accessed instances. Horizontal
fragments are built by selection. The original relation is
reconstructed by fragment union. A variant, derived horizontal
fragmentation, consists in partitioning a relation with respect to
predicates defined on another relation.

Finally, hybrid fragmentation consists of either horizontal
fragments that are subsequently vertically fragmented, or vertical
fragments that are subsequently horizontally fragmented.

\subsection{Data Warehouse Fragmentation}
\label{sec:relational-fragmentation}

Many research studies address the issue of fragmenting relational
data warehouses either to efficiently process analytical queries or
to distribute the warehouse.

To improve ad-hoc query performance, Datta \textit{et al.} exploit a
vertical fragmentation of facts to build the Cuio index~\cite{DattaRT99}, while Golfarelli \textit{et al.} apply the same
fragmentation on warehouse views \cite{GolfarelliMR99}.
Munneke \textit{et al.} propose a fragmentation methodology for a
multidimensional database~\cite{MunnekeWM99}. Fragmentation consists
in deriving a global data cube into fragments containing a subset of
data. This process is defined by the slice and dice operation. The
authors also define another fragmentation strategy, server, that
removes one or several dimensions from a hypercube to produce
fragments with fewer dimensions than the original data cube.
Bellatreche and Boukhalfa apply horizontal fragmentation to a
star-schema \cite{BellatrecheB05}. Their fragmentation strategy is
based on a query workload and exploits a genetic algorithm to select
a partitioning schema. This algorithm aims at choosing an optimal
fragmentation schema that minimizes query cost.
Finally, Wu and Buchmaan recommend to combine horizontal and
vertical fragmentation for query optimization \cite{WuB97}. A fact
table can be horizontally partitioned with respect to one or more
dimensions. It can also be vertically partitioned according to its
dimensions, i.e., all the foreign keys to the dimension tables are
partitioned as separate tables.

To distribute a data warehouse, Noaman \textit{et al.} exploit a
top-down strategy that uses horizontal fragmentation
\cite{NoamanB99}. The authors propose an algorithm for deriving
horizontal fragments from the fact table based on queries that are
defined on all dimension tables.
Finally, Wehrle \textit{et al.} propose to distribute and query a
warehouse on a computing grid~\cite{WehrleMT05}. They use derived
horizontal fragmentation to split the data warehouse and build a
so-called \textit{block of chunks}, a data set defining a fragment.

In summary, these proposals generally exploit static derived
horizontal fragmentation to reduce irrelevant data access rate and
efficiently process join operations across multiple relations
\cite{BellatrecheB05,NoamanB99,WehrleMT05}. In the literature, the
prevalent methods used for derived horizontal fragmentation are the
following~\cite{KoreichiB97}.

\begin{itemize}

\item \textbf{Predicate construction}. This method fragments a relation by using a complete and minimal set
of predicates \cite{NoamanB99}. Completeness means that two relation
instances belonging to the same fragment have the same probability
of being accessed by any query. Minimality garantees that there is
no redundancy in predicates.

\item \textbf{Affinity-based fragmentation}.
This method is an adaptation of vertical fragmentation methods to
horizontal fragmentation \cite{NavatheKR95}. It is based on the
predicate affinity concept \cite{ZhangO94}, where affinity defines
query frequency. Specific matrices (predicate usage and affinity
matrices) are exploited to cluster selection predicates. A cluster
is defined as a selection predicate cycle and forms a dimension
graph fragment.

\end{itemize}

\subsection{XML Database Fragmentation}
\label{sec:xml-fragmentation}

Recently, several fragmentation techniques for XML data have been
proposed. They split an XML document into a new set of XML
documents. Their main objective is either to improve XML query
performance \cite{BonifatiC07,GertzB03,MaSHK03} or to distribute or
exchange XML data over a network \cite{BonifatiMCJ04,BoseF05}.

To fragment XML documents, Ma \textit{et al.} define a new
fragmentation type: \textit{split} \cite{MaS03,MaSHK03}, which is
inspired from the oriented-object domain. This fragmentation splits
XML document elements and assigns a reference to each sub-element.
The references are then added to the Document Type Definition (DTD)
defining the XML document. The authors extend the DTD and XML-QL
languages.
Bonifati \textit{et al.} also propose a fragmentation strategy for
XML documents that is driven by structural constraints
\cite{BonifatiC07,BonifatiCZ06}. This strategy uses both heuristics
and statistics.
Andrade \textit{et al.} propose to apply fragmentation to an
homogeneous XML collection~\cite{AndradeRBBM06}. They adapt
traditional fragmentation techniques to an XML document collection
and base their proposal on the Tree Logical Class (TLC) algebra
\cite{PaparizosWLJ04}. The authors also evaluate these techniques
and show that horizontal fragmentation provides the best
performance. 

Gertz and Bremer introduce a distribution approach for an XML
repository \cite{GertzB03}. They propose a fragmentation method and
outline an allocation model for distributed XML fragments in a
centralized archirecture. Gertz and Bremer also define horizontal
and vertical fragmentation for an XML document. A fragment is
defined with a path expression language, called \textit{XF}, which
is derived from XPath. This fragment is obtained by applying an
\textit{XF} expression on a graph \textit{RG} representing XML data.
Moreover, the authors define exclusion expressions that ensure
fragment coherence and disjunction.

Bose and Fegaras use XML fragments for data exchange in a
peer-to-peer network (P2P), called XP2P \cite{BoseF05}. XML
fragments are interrelated and each is uniquely identified by an
\textit{ID}. The authors propose a fragmentation schema, called
\textit{Tag Structure}, to define the structure of data and
fragmentation information.
Bonifati \textit{et al.} also define XML fragments for a P2P
framework \cite{BonifatiMCJ04}. An XML fragment is obtained and
identified by a single path expression, a root-to-node path
expression \textit{XP}, and managed on a specific peer. In addition,
the authors associate to each fragment two path expressions:
\textit{super fragment} and \textit{child fragment}. These paths
ensure the identification of fragments and relationships.

In summary, these proposals adapt classical static fragmentation
methods to split XML data. An XML fragment is defined and identified
by a path expression \cite{BonifatiMCJ04} or an XML algebra operator
\cite{AndradeRBBM06}. Fragmentation is performed on a single XML
document \cite{MaS03,MaSHK03} or on an homogeneous XML collection
\cite{AndradeRBBM06}. Note that XML data warehouse fragmentation has
not been addressed yet, to the best of our knowledge.

\subsection{Data Mining-based Fragmentation}
\label{sec:DataMiningBasedFragmentation}

Although data mining has proved useful for selecting physical data structures that enhance performance, such as indexes or materialized views \cite{adbis06,innovations07}, few approaches exploit data mining for fragmentation.

Gorla and Betty exploit
association rules (by adapting the Apriori
algorithm \cite{AgrawalS94}) for vertical fragmentation approach in relational
databases \cite{GorlaJ05}.

Darabant and Campan propose the horizontal fragmentation method for
object-oriented distributed databases based on k-means clustering
we inspire from \cite{Darabant04}. This method clusters object instances
into fragments by taking all complex relationships
between classes into account (aggregation, associations and links induced by
complex methods).

Finally, Fiolet and Toursel propose a parallel, progressive clustering
algorithm to fragment a
database and distribute it over a grid \cite{FioletT05bis}. It is inspired by the CLIQUE sequential
clustering algorithm that consists in clustering data by projection.

Though in limited number, these studies clearly demonstrate how data
mining can be used for vertical and horizontal fragmentation,
through association rule mining and clustering, respectively. They
are also static, though.

\section{K-Means-based Fragmentation}
\label{sec:KMeansBasedFragmentation}

Although XML data warehouse architectures from the literature share a lot of concepts (mostly originating from classical data warehousing), they are
nonetheless all different. Hence, we proposed a unified, reference XML data warehouse model that synthesizes and enhances existing models~\cite{edwm208}, and on which we base our fragmentation work. We first recall this model before detailing our fragmentation approach.

\subsection{XML Warehouse Reference Model}
\label{sec:XMLWarehouseReferenceModel}

XML warehousing approaches assume that the warehouse is composed of XML documents that represent both facts and dimensions. All these studies mostly differ in the way dimensions are handled and the number of XML documents that are used to store facts and dimensions. A performance evaluation study of these different representations showed that representing facts in one single XML document and each dimension in one XML document allowed the best performance \cite{DoulkiliBB06}. Moreover, this representation also allows to model constellation schemas without duplicating dimension information. Several fact documents can indeed share the same dimensions. Hence, we adopt this architecture model. More precisely, our reference data warehouse is composed of the following XML documents:
\begin{itemize}
\item \textit{dw-model.xml} that represents warehouse metadata;
\item a set of $facts_f.xml$ documents that each store information related
to set of facts $f$;
\item a set of $dimension_{d}.xml$ documents that each store a given
dimension $d$'s member values.
\end{itemize}

\begin{figure}[hbt]
\centering
\epsfig{file=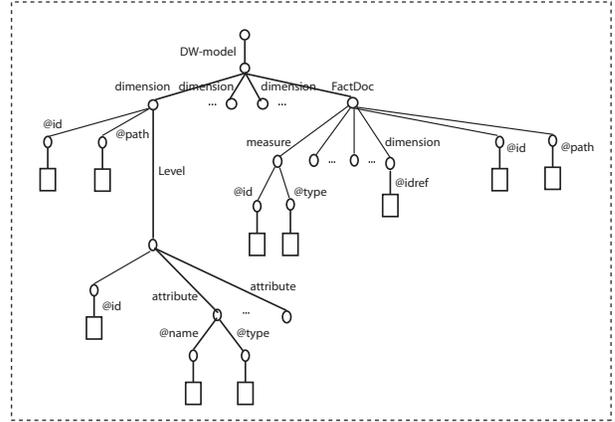, width=8cm}
\caption{\textit{dw-model.xml} graph structure}
\label{fig:dw-model}
\end{figure}

The \textit{dw-model.xml} document (Figure~\ref{fig:dw-model}) defines the multidimensional
structure of the warehouse. Its root node, \textit{DW-model}, is
composed of two types of nodes: \textit{dimension} and
\textit{FactDoc}. A \textit{dimension} node defines one
dimension, its possible hierarchical levels ($Level$ elements) and
attributes (including their types), as well as the path to the
corresponding $dimension_{d}.xml$ document. A \textit{FactDoc}
element defines a fact, i.e., its measures, references
to the corresponding dimensions, and the
path to the corresponding $facts_f.xml$ document.

\begin{figure}[hbt]
\centering
\epsfig{file=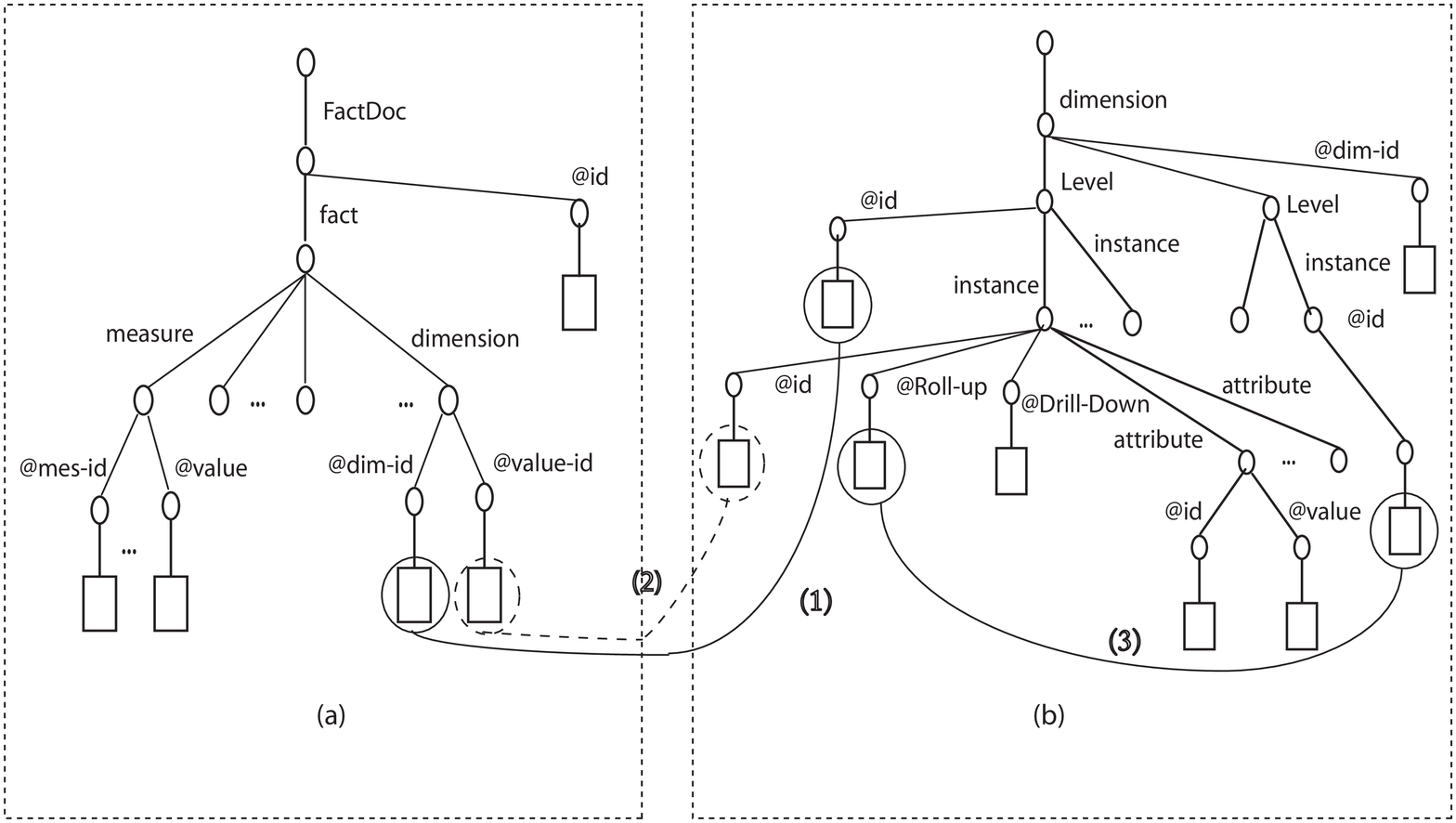, width=8cm}
\caption{$facts_f.xml$ (a) and $dimension_{d}.xml$ (b) graph structures}
\label{fig:fact-dimension}
\end{figure}

A $facts_f.xml$ document stores facts (Figure~\ref{fig:fact-dimension}(a)). The document root node,
\textit{FactDoc}, is composed of \textit{fact} subelements that each
instantiate a fact, i.e., measure values and dimension references. These identifier-based references support the
fact-to-dimension relationship.

Finally, a $dimension_{d}.xml$ document helps instantiate one dimension,
including any hierarchical level (Figure~\ref{fig:fact-dimension}(b)). Its root node, \textit{dimension},
is composed of \textit{Level} nodes. Each one defines a hierarchy
level composed of \textit{instance} nodes that each define the
level's member attribute values. In addition, an \textit{instance}
element contains \textit{Roll-up} and \textit{Drill-Down} attributes
that define the hierarchical relationship within dimension $d$.

\pagebreak

\subsection{Fragmentation Approach}
\label{sec:FragmentationApproach}

\subsubsection{Principle}
\label{sec:Principle}

\begin{figure}[hbt]
\centering
\epsfig{file=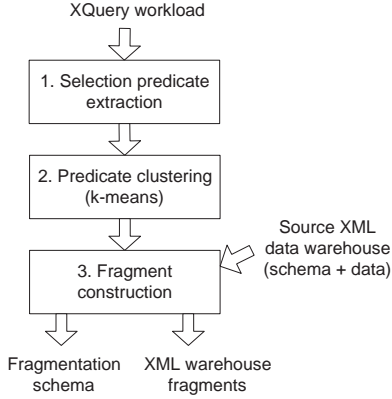, width=6cm}
\caption{K-means-based fragmentation principle}
\label{fig:principle}
\end{figure}

Since the aim of fragmentation is to optimize query response time, the prevalent fragmentation strategies are workload driven \cite{BellatrecheB05,BonifatiC07,GertzB03,NavatheKR95,NoamanB99}. More precisely, they exploit selection predicates found in workload queries to derive suitable fragments. Our approach also belongs to this family. Its general principle is summarized in Figure~\ref{fig:principle}. It outputs both a fragmentation schema (metadata) and the fragmented XML warehouse. It is subdivided into three steps that are detailed in the following sections:
\begin{enumerate}
    \item selection predicate extraction from the query workload;
    \item predicate clustering with the k-means method;
    \item fragment construction with respect to predicate clusters. 
\end{enumerate}

\vspace{-0.25cm}

\subsubsection{Selection Predicate Extraction}
\label{sec:SelectionPredicateExtraction}

Selection predicate set $P$ is simply parsed from workload $W$. For example, let $W_S$ be the sample XQuery workload provided in Figure~\ref{fig:req} and $P_S$ the corresponding predicate set. $P_S = \{p_1, p_2, p_3, p_4, ...\}$, where:\\
$p_1 = \$y/attribute[$@$id="c\_nation\_key"]/@value$>$"15"$,\\
$p_2 = \$y/attribute[$@$id="c\_nation\_key"]/$@$value="13"$,\\
$p_3 = \$y/attribute[$@$id="p\_type"]/$@$value="PBC"$ and\\
$p_4 = \$y/attribute[$@$id="d\_date\_name"]/@value="Sat."$.
For example, $p_2$ and $p_3$ are selection predicates obtained from query $q_{2} \in W_S$.

\begin{figure}[hbt]
\centering{ \scriptsize{
\begin{tabular}{cl}
$q_{1}$ &   for \$x in //FactDoc/Fact,\\
&           \$y in //dimension[$@$dim-id="Customer"]/Level/instance\\
&            where \$y/attribute[$@$id="c\_nation\_key"]/@value$>$"15"\\
&          and \$x/dimension[$@$dim-id="Customer"]/$@$value-id=\$y/$@$id\\
&       return \$x\\ \\
$q_{2}$ &   for \$x in //FactDoc/Fact,\\
&           \$y in //dimension[$@$dim-id="Customer"]/Level/instance,\\
&           \$z in //dimension[$@$dim-id="Part"]/Level/instance\\
&            where \$y/attribute[$@$id="c\_nation\_key"]/@value="13"\\
&            and \$y/attribute[$@$id="p\_type"]/@value="PBC"\\
&          and \$x/dimension[$@$dim-id="Customer"]/$@$value-id=\$y/$@$id\\
&          and \$x/dimension[$@$dim-id="Part"]/$@$value-id=\$z/$@$id\\
&       return \$x\\
\dots \\
$q_{10}$ &   for \$x in //FactDoc/Fact,\\
&           \$y in //dimension[$@$dim-id="Customer"]/Level/instance,\\
&           \$z in //dimension[$@$dim-id="Date"]/Level/instance\\
&            where \$y/attribute[$@$id="c\_nation\_key"]/@value="13"\\
&            and \$y/attribute[$@$id="d\_date\_name"]/@value="Sat."\\
&          and \$x/dimension[$@$dim-id="Customer"]/$@$value-id=\$y/$@$id\\
&          and \$x/dimension[$@$dim-id="Part"]/$@$value-id=\$z/$@$id\\
&       return \$x\\
\end{tabular}}
} \caption{XQuery workload snapshot $W_S$}\label{fig:req}
\end{figure}

Parsed predicates are then coded in a query-predicate matrix $QP$ whose general term $QP_{ij}$ equals to 1 if predicate $p_{j} \in P$ appears in query $q_{i} \in W$, and to 0 otherwise. For example, the $QP_S$ matrix corresponding to $W_S$ and $P_S$ is featured in Table~\ref{Table:qps}.

\begin{table}[hbt]
\begin{center}
\begin{tabular}{|l|c|c|c|c|c|}
\hline
& \small{$p_1$} & \small{$p_2$} & \small{$p_3$} & \small{$p_4$} &  \small{...} \\
\hline
\small{$q_1$} & \small{1} & \small{0} & \small{0} & \small{0} & \\
\hline
\small{$q_2$} & \small{0} & \small{1} & \small{1} & \small{0} & \\
\hline
\small{...} & & & & & \\
\hline
\small{$q_{10}$} & \small{0} & \small{0} & \small{1} & \small{1} & \\
\hline
\end{tabular}
\caption{Sample query-predicate matrix $QP_S$}
\label{Table:qps}
\end{center}
\end{table}

\subsubsection{Predicate Clustering}
\label{sec:SelectionPredicateClustering}

Our objective is to derive fragments that optimize data access for a given set of queries. Since horizontal fragments are built from predicates, clustering predicates with respect to queries achieves our goal. Intuitively, we ideally seek to build rectangles (clusters) of 1s in matrix $QP$. We chose the widely-used k-means algorithm \cite{kmeans} for clustering. This algorithm inputs vectors of object attributes (columns of $QP$ in our case). It attempts to find the centers of natural clusters in source data by minimizing total intra-cluster variance $\sum_{i=1}^k \sum_{x_j \in C_i} (x_j - \mu_i)^2$, where $C_i, i=1, ..., k$ are the $k$ output clusters and $\mu_i$ is the centroid (mean point) of points $x_j \in C_i$. Let $C$ be the set of all clusters $C_i$.

Usually, having $k$ as an input parameter is viewed as a drawback in clustering. In our case, this turns out to be an advantage, since we want to limit the number of clusters/fragments, typically to the number of nodes the XML data warehouse will be distributed on.

In practice, we used the Weka \cite{weka} SimpleKMeans implementation of k-means. SimpleKMeans uses the Euclidean distance to compute distances between points and clusters. It directly inputs matrix $QP$ (acually, the $p_j$ vectors) and $k$, and outputs set of predicate clusters $C$. For example, on matrix $QP_S$ with $k = 2$, SimpleKMeans outputs:
\begin{center}
$C_S = \{\{p_1\}, \{p_2, p_3, p_4\}\}$.
\end{center}

\subsubsection{Fragment Construction}
\label{sec:FragmentConstruction}

\begin{figure}[hbt]
\centering
\epsfig{file=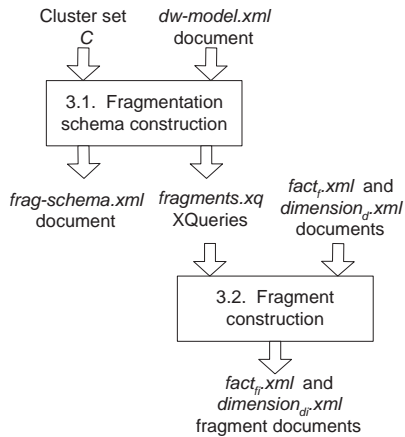, width=6.1cm}
\caption{Fragment construction substeps}
\label{fig:fragments}
\end{figure}

\begin{figure}[hbt]
\centering
\epsfig{file=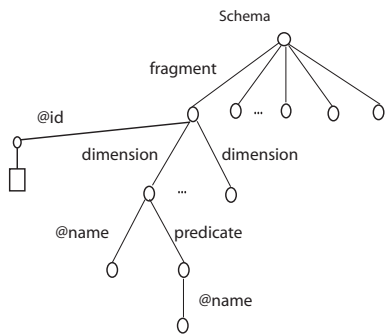, width=5cm}
\caption{\emph{frag-schema.xml} graph structure}
\label{fig:frag-schema}
\end{figure}

The fragmentation construction step is itself subdivided into two substeps (Figure~\ref{fig:fragments}). First, predicate cluster set $C$ is joined to warehouse schema (document \textit{dw-model.xml}) to produce an XML document named \emph{frag-schema.xml} that represents the fragmentation
schema (Figure~\ref{fig:frag-schema}). Its root
node, \emph{Schema}, is composed of \emph{fragment} elements. Each fragment is identified by an
\emph{@id} attribute and contains \emph{dimension} elements. A
dimension element is identified by a \emph{@name} attribute and
contains \emph{predicate}
elements that store the predicates used for fragmentation. For example, the fragmentation schema \emph{frag-schema$_S$.xml} corresponding to cluster set $C_S$ is provided in Figure~\ref{fig:frag-doc}.

\begin{figure}[hbt]
\centering{
\scriptsize{
\begin{tabular}{p{6.5cm}}
<Schema>\\
    \hspace*{1cm} <fragment id="f1">\\
        \hspace*{2cm} <dimension name="Customer">\\
            \hspace*{3cm} <predicate name="p1" />\\
        \hspace*{2cm} </dimension>\\
    \hspace*{1cm} </fragment>\\
    \hspace*{1cm} <fragment id="f2">\\
        \hspace*{2cm} <dimension name="Customer">\\
            \hspace*{3cm} <predicate name="p2" />\\
        \hspace*{2cm} </dimension>\\
        \hspace*{2cm} <dimension name="Part">\\
            \hspace*{3cm} <predicate name="p3" />\\
        \hspace*{2cm} </dimension>\\
        \hspace*{2cm} <dimension name="Date">\\
            \hspace*{3cm} <predicate name="p4" />\\
        \hspace*{2cm} </dimension>\\
    \hspace*{1cm} </fragment>\\
</Schema>
\end{tabular}}
}
\caption{Sample \emph{frag-schema$_S$.xml} document}
\label{fig:frag-doc}
\end{figure}

In this process, we also output a set of XQueries (the \emph{fragments.xq} script) that, applied to the XML data warehouse (i.e., the whole set of $facts_f.xml$ and $dimension_d.xml$ documents), produces the actual fragments, which we store in a set of $facts_{f_i}.xml$ and $dimension_{d_i}.xml$ documents, $i=1, ..., k+1$. As fragments, these documents indeed bear the same schema than the original warehouse. The $(k+1)^{th}$ fragment is based on an additional predicate, denoted $ELSE$, which is the negation of the conjunction of all predicates in $P$ and is necessary to ensure fragmentation completeness (Section~\ref{sec:relational-fragmentation}). In our running example, $ELSE = \neg (p_1 \wedge  p_2 \wedge  p_3 \wedge  p_4)$.

Figure~\ref{fig:frag-q} provides an excerpt from the \emph{fragments$_{S}$.xq} script that helps build fragment $f2$ from Figure~\ref{fig:frag-doc}. Dimension fragments are generated first, one by one, through selections exploiting the predicate(s) associated to the current dimension (three first queries from Figure~\ref{fig:frag-q}). Then, fragmentation is derived on facts by joining the original fact document to the newly-created dimension fragments (last query).

\begin{figure}[hbt]
\centering{
\scriptsize{
\begin{tabular}{p{8cm}}
element dimension\{ attribute dim-id\{Customer\}, element Level\{ \\
attribute id \{Customers\},\\
for \$x in document("$dimension_{Customer}.xml$")//Level \\
where \$x//attribute[$@$id="c\_nation\_key"]/$@$value="13"] \\
return \$x \} \\
\}\\
element dimension\{ attribute dim-id\{Part\}, element Level\{ \\
attribute id \{Part\},\\
for \$x in document("$dimension_{Part}.xml$")//Level \\
where \$x//attribute[$@$id="p\_type"]/$@$value="PBC"] \\
return \$x \} \\
\}\\
element dimension\{ attribute dim-id\{Date\}, element Level\{ \\
attribute id \{Date\},\\
for \$x in document("$dimension_{Date}.xml$")//Level \\
where \$x//attribute[$@$id="d\_date\_name"]/$@$value="Sat."] \\
return \$x \} \\
\}\\
element FactDoc \{\\
for \$x in //FactDoc/Fact,\\
    \$y in document("$dimension_{Customer_{f2}}.xml$")//instance,\\
    \$z in document("$dimension_{Part_{f2}}.xml$")//instance,\\
    \$t in document("$dimension_{Date_{f2}}.xml$")//instance\\
where \$x/dimension[$@$dim-id="Customer"]/$@$value-id=\$y/$@$id\\
and \$x/dimension[$@$dim-id="Part"]/$@$value-id=\$z/$@$id\\
and \$x/dimension[$@$dim-id="Date"]/$@$value-id=\$t/$@$id\\
return \$x \\
\}
\end{tabular}}
}
\caption{Excerpt from sample \emph{fragments$_{S}$.xq} script}
\label{fig:frag-q}
\end{figure}

\vspace{-0.15cm}

\section{Experiments}
\label{sec:ExperimentalPerformanceStudy}

Since derived horizontal fragmentation is a NP-hard problem \cite{Boukhalfa08BR} solved by heuristics, we choose to validate our proposal experimentally.

\subsection{Experimental conditions}

We use XWeB (XML Data Warehouse Benchmark) \cite{MahboubiD06} as a test platform. XWeB is based on
the reference model defined in Section \ref{sec:XMLWarehouseReferenceModel}, and proposes
a test XML data warehouse and its associated XQuery decision-support
workload.

XWeB's warehouse consists of \emph{sale} facts characterized by the
amount (of purchased products) and quantity (of purchased products)
measures. These facts are stored in the $facts_{sales}.xml$ document
and are described by four dimensions: \emph{Customer},
\emph{Supplier}, \emph{Date} and \emph{Part} stored in the
$dimension_{Customer}.xml$, $dimension_{Supplier}.xml$,\\
$dimension_{Date}.xml$ and $dimension_{Part}.xml$ documents,
respectively. XWeB's warehouse characteristics are displayed in
Table \ref{Table:dw-characteristics}.

\begin{table}[hbt]
\begin{center}
\begin{tabular}{|l|l|}
\hline \small{\textbf{Facts}} & \small{\textbf{Maximum number of cells}}\\
\hline Sale facts & 7000 \\
\hline
\hline \small{\textbf{Dimensions}} & \small{\textbf{Number of instances}}\\
\hline Customer & 1000 \\
\hline Supplier & 1000 \\
\hline Date  & 500 \\
\hline Part  & 1000 \\
\hline
\hline \small{\textbf{Documents}} & \small{\textbf{Size (MB)}}\\
\hline $facts_{sales}.xml$ & 2.14\\
\hline $dimension_{Customer}.xml$ & 0.431\\
\hline $dimension_{Supplier}.xml$ & 0.485\\
\hline $dimension_{Date}.xml$  & 0.104\\
\hline $dimension_{Part}.xml$  & 0.388\\
\hline
\end{tabular}
\caption{XWeB warehouse characteristics}
\label{Table:dw-characteristics}
\end{center}
\end{table}

XWeB's workload is composed of queries that exploit the warehouse
through join and selection operations. We extend this workload by
adding queries and selection predicates in order to obtain a
significant fragmentation. Due to space constraints, our workload is only available on-line\footnote{\scriptsize{http://eric.univ-lyon2.fr/$\sim$hmahboubi/Workload/workload.xq}}.

We ran our tests on a Pentium 2 GHz PC with 1 GB of main memory and
an IDE hard drive under Windows XP. We use the X-Hive XML native
DBMS\footnote{\scriptsize{http://www.x-hive.com/products/db/}} 
to store and query the warehouse. Our code is written in Java and connects to X-Hive and Weka through their respective Application Programming Interfaces (APIs). It is available on demand.

\subsection{Fragmentation Strategy Comparison}
\label{sec:FragmentationStrategyComparison}

In this first series of experiments, we aim at comparing our k-means-based horizontal fragmentation approach (denoted KM) to the classical derived horizontal fragmentation techniques, namely by predicate construction (PC) and affinity-based (AB) primary fragmentation (Section~\ref{sec:relational-fragmentation}), which we adapted to XML data warehouses \cite{dapd}. We also record performance when no fragmentation is applied (NF), for reference.

\subsubsection{Query Response Time}
\label{sec:QueryResponseTime}

This experiment measures workload execution time with the three fragmentation strategies we adopted. For KM, we arbitrarily fixed $k=8$, which could correspond to a computer cluster's size. The fragments we achieve are stored in distinct
collections to simulate data distribution. Each collection can
indeed be considered to be stored on a distinct node and can be identified, targeted and queried separately. To measure query execution time
over a fragmented warehouse, we first identify the required
fragments with the \emph{frag-schema.xml} document. Then, we execute the query over
each fragment and save execution time. To simulate parallel
execution, we only consider maximum execution time.

\begin{figure}[hbt]
\centering
\epsfig{file=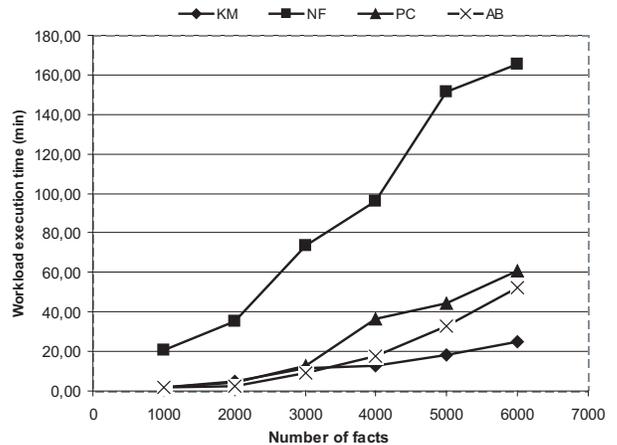, width=9.5cm}
\caption{Fragmentation efficiency comparison}
\label{fig:exp11}
\end{figure}

Figure~\ref{fig:exp11} plots workload response time with respect to data warehouse size (expressed in number of facts). It clearly shows that fragmentation significantly improves response time, and that KM fragmentation performs better than PC and AB fragmentation when the warehouse scales up. Workload execution time is indeed, on an average, 86.5\% faster with KM fragmentation than with NF, and 36.7\% faster with KM than with than AB. We believe our approach performs better than classical derived horizontal fragmentation techniques because these latter produce many more fragments (159 with PC and 119 with AB \emph{vs.} 9 with KM). Hence, at workload execution time, queries must access many fragments (up to 50 from our observations), which multiplies query distribution and result reconstruction costs. The number of accessed fragments is much lower with KM (typically 2 fragments in our experiments).

\pagebreak

\subsubsection{Fragmentation Overhead}
\label{sec:FragmentationOverhead}

We also compare the PC, AB and KM ($k=8$) fragmentation strategies in terms of overhead (i.e., fragmentation algorithm execution time). When assessing performance, it is indeed necessary to find a fair trade-off between gain and overhead. Table~\ref{Table:exp12} summarizes the results we obtain for an arbitrarily fixed data warehouse size of 3,000 facts. It shows that KM clearly outperforms AB and PC, which is in line with these algorithms' complexities~: $O(|P|)$, $O(|P|^2)$ and $O(2^{|P|})$, respectively. While AB and PC would have to run off-line, KM could on the other hand be envisaged to run on-line.

\begin{table}[hbt]
\begin{center}
\begin{tabular}{|l|c|c|c|}
\hline
& \small{\textbf{PC}} & \small{\textbf{AB}} & \small{\textbf{KM}} \\
\hline
\small{\textbf{Execution time (h)}} & \small{16.8} & \small{11.9} & \small{0.25} \\
\hline
\end{tabular}
\caption{Fragmentation overhead comparison}
\label{Table:exp12}
\end{center}
\end{table}

\subsection{Influence of Number of Clusters}
\label{sec:InfluenceOfNumberOfClusters}

\begin{figure}[hbt]
\centering
\epsfig{file=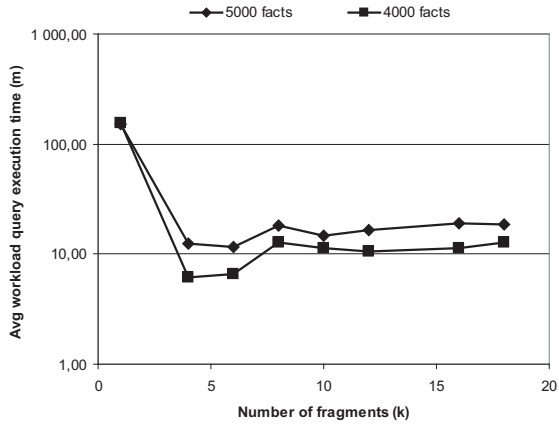, width=8cm}
\caption{Influence of number of clusters}
\label{fig:exp2}
\end{figure}

In this experiment, we fixed data warehouse size (to 4,000 and 5,000 facts, respectively) and varied KM parameter $k$ to observe its influence on workload response time. Figure~\ref{fig:exp2} confirms that performance improves quickly when fragmentation is applied, but tends to degrade when the number of fragments increases, as we explained in Section~\ref{sec:QueryResponseTime}. Furthermore, it hints that an optimal number of clusters \emph{for our test data warehouse and workload} lies between 4 and 6, making us conclude that over-fragmentation must be detected and avoided. Note that, on Figure~\ref{fig:exp2}, $k=1$ corresponds to the NF experiment (this one fragment is the original warehouse).

\section{Conclusion}
\label{sec:Conclusion}

In this paper, we have introduced an approach for fragmenting XML data warehouses that is based on data mining, and more precisely clustering and the k-means algorithm. Classical derived horizontal fragmentation strategies run automatically and output an unpredictable number of fragments, which is nonetheless crucial to keep under control. By contrast, our approach allows to fully master the number of fragments through the k-means $k$ parameter.

To validate our proposal, we have compared our fragmentation strategy to XML adaptations of the two prevalent fragmentation methods for relational data warehouses. Our experimental results show that our approach, by producing a lower number of fragments, outperforms both the others in terms of performance gain and overhead.

Now that we have efficiently fragmented an XML data warehouse, our more direct perspective is to distribute it on a data grid. This raises several issues that include processing a global query into subqueries to be sent to the right nodes in the grid, and reconstructing a global result from subquery
results. Properly indexing the distributed warehouse to guarantee good performance shall also be very important.

Finally, in a continuous effort to minimize the data warehouse administration function and aim at autoadministrative systems \cite{adbis06,innovations07}, we plan to make our data mining based-fragmentation strategy dynamic. The idea is to perform incremental fragmentation when the warehouse is refreshed. This could be achieved with the help of an incremental variant of the k-means algorithm \cite{inc_kmeans}.

\section{Acknowledgments}

The authors would like to thank Houssem Aissa, Anouar Benzakour, Kevin du Repaire and Hamza El Kartite, who participated in coding our approach in Java.

\bibliographystyle{abbrv}
\bibliography{fragmentation}  

\end{document}